\documentstyle[twocolumn,epsf,aps]{revtex}

\newcommand{\GRLower}[1]{\smash{\lower 0.25ex \hbox{#1}}}

\def\goes{\rightarrow}
 
\def\SM{SU(2)_{\hbox{\GRLower{${\scriptstyle L}$}}} \times
        U(1)_{\hbox{\GRLower{${\scriptstyle Y}$}}}}
      \def\mz{M_{Z}}
      
      \def\mq{m_{q}}

                 \def\mmu{\mid \!\! \mu \!\! \mid}
                 \def\tb{\tan \beta}
                 \def\mg{M_{\widetilde g}}
                 \def\msq{m_{\widetilde q}}
                 \def\msql{m_{{\widetilde q}_L}}
                 \def\msqr{m_{{\widetilde q}_R}}
                 \def\mstl{m_{{\widetilde t}_L}}
                 \def\mstr{m_{{\widetilde t}_R}}
                 \def\msbl{m_{{\widetilde b}_L}}
                 \def\msbr{m_{{\widetilde b}_R}}
                 \def\mssl{m_{{\widetilde{{\scriptstyle s}}_L}}}
                 \def\msdl{m_{{\widetilde d}_L}}

                 \def\aq{A_{\widetilde q}}

                 \def\mh{M_{h}}
                 
                 \def\mA{M_{A}}

      \def\h{h^0}
      
      \def\A{A^0}
\begin{document}
\draft
\title{Combined Squark and Gluino Mass Bounds from LEP Data}
\author{Mike Bisset\cite{email},
Sreerup Raychaudhuri\cite{email}, }
\address{Theoretical Physics Group, Tata Institute of
Fundamental Research,Mumbai 400 005, India}
\author{ and Dilip Kumar Ghosh\cite{email}}
\address{Department of Physics, Univ. of Bombay,
Vidyanagari, Santa Cruz, Mumbai 400 098, India}
\date{\today}
\maketitle

\begin{abstract}
Under the assumption of gaugino mass unification at a high scale,
chargino and neutralino masses depend on the value of the gluino mass, 
which itself becomes a function of squark masses through self-energy 
corrections. We demonstrate that this leads to combined bounds on squark 
and gluino masses from the limits on chargino, neutralino and Higgs 
boson masses obtained in the CERN LEP-1 and LEP-1.5 runs. These bounds 
turn out to be comparable to those obtained from direct searches at 
the Fermilab Tevatron and may be expected to improve as LEP energies 
go higher. 
\end{abstract}



Searches for supersymmetric partners of known particles (sparticles)
are high priority items at particle accelerators. 
The non-discovery of such sparticles 
constrains supersymmetric models, such as the popular
Minimal Supersymmetric Standard Model (MSSM) 
(for reviews, see \cite{SUSY1,SUSY2,SUSY3}).
The direct searches for strongly-interacting 
sparticles --- squarks and gluinos --- in $p \bar p$ collisions at the 
Fermilab Tevatron have led to well-known bounds on the masses of these 
sparticles.   Direct as well as indirect searches for electroweak 
sparticles --- including charginos and neutralinos --- in $e^+ e^-$ 
collisions at the CERN LEP collider have also yielded 
constraints on the MSSM parameter space.  However, the links 
between these two classes of sparticles have not been fully explored in 
previous analyses \cite{DiazKing}.

In this letter, we point out that bounds on chargino and neutralino (and 
Higgs boson $\h$) masses from the LEP data can be translated into 
bounds on the squark-gluino mass plane similar to those 
 obtained from the direct Tevatron searches.
The crucial features of this analysis are ($a$) the 
assumption of gaugino mass unification at a high energy 
scale \cite{GUTrel}, which relates the 
gluino mass $\mg$ to the soft supersymmetry(SUSY)-breaking parameters 
$M_1,M_2$ 
in the $\SM$ sector, and ($b$) the observation that the gluino mass (and 
hence $M_1,M_2$) which determines the chargino and neutralino masses at 
the different energy scales explored by the LEP experiments is a 
{\it running} mass driven by squark loops which differs significantly 
(by $\sim 50$--$100\, \hbox{GeV}$) from the physical mass probed at 
the Tevatron.

Incorporating gaugino mass unification, chargino and 
neutralino mass-matrices depend upon the three parameters: 
the `gluino mass', $\mu$ and $\tb$.  (Here $\mu$ is the 
Higgsino-mixing parameter and $\tb$ is the ratio of vacuum expectation 
values of the two scalar doublets in the theory.) 
However, the `gluino mass' parameter here is actually the
gluino mass $\mg(\sqrt{s})$ evaluated at the LEP energy-scale 
$\sqrt{s}$ \cite{scale}.
It is a function of both the physical gluino mass $\mg$ and the squark 
masses and couplings (through radiative corrections) \cite{MVK}. 
Thus chargino and neutralino 
masses and couplings should be considered functions of $\mg,\mu,\tb$ as well 
as the mass-parameters of the squark sector. In principle, this 
brings into play the full set of inputs which go into the construction of 
the squark mass-squared matrices \cite{sqkmat,MikeT}; {\it i.e.}, 
the soft SUSY-breaking masses $\msql,\msqr$
of left and right squarks respectively and $\aq$, the trilinear squark 
coupling, {\it for each flavor}.
This means that the parameter space that should be considered when 
determining constraints from LEP data must be expanded from the 
traditional $\mg,\mu,\tb$-parameter set to incorporate many
new independent inputs from the squark sector.

The above proposition is rather cumbersome, so all the soft 
SUSY-breaking squark masses at the weak scale
will be set to a common value, $\msq$, 
and all the trilinear couplings $\aq$ set to zero. 
This is employed in the Tevatron analyses\cite{CDF,D0}, and we will follow 
their example here, in part to
facilitate comparison of our LEP constraints with those from the the
Fermilab Tevatron -- the main thrust of this letter. 
We will comment on the 
effects of relaxing these assumptions when appropriate.
The entire squark sector is thus represented by the single parameter 
$\msq$. 

The LEP-1 experimental constraints imposed on the MSSM 
(with the additional assumptions described above) are the 
following \cite{LEP1rev}:
\newline
{\bf 1.} The sparticle 
contribution to the $Z^0$ boson width. This must be less
than the difference between the Standard Model (SM) prediction and the
experimental value at 95\% CL --- roughly 
$23.1\, \hbox{MeV}$. 
\newline
{\bf 2.}  The partial decay width of the $Z^0$ boson to a pair of lightest
neutralinos (LSP's).  This
contributes to the invisible width, and thus must be
less than the difference between the experimental number at 95\% CL and
the SM prediction with 3 neutrino generations --- 
about $8.4\, \hbox{MeV}$.
\newline
{\bf 3.} The branching ratio of the $Z^0$ boson to any pair of dissimilar
neutralinos.  Direct LEP searches for such event topologies among the
millions of $Z^0$-decays tallied thus far restrict this to be less that 
$10^{-5}$ \cite{degW1LSP}.
\newline
{\bf 4.} The physical masses of all the squarks.  
All charged sfermions must have masses
larger than $\mz/2$.
Sleptons do not directly 
enter into our analysis and so we set their masses to be very heavy.
\newline
{\bf 5.}  The combined masses of the 
$CP$-odd pseudoscalar and $CP$-even lighter scalar Higgs bosons.
The decay channel $Z^0 \goes \h \A$ is strongly constrained by
LEP searches basically requiring that $\mA + \mh$ $>$ $M_Z$.
\newline
{\bf 6.}  The partial width for the Bjorken process with the lighter 
scalar Higgs boson $\h$. This should not exceed the corresponding 
partial width for the $Z^0$-decay to a SM Higgs boson, where the mass of 
this SM Higgs boson is given by the experimental bound of 
$65.2\, \hbox{GeV}$ \cite{Hllim}. 

\parindent 0pt
We also impose one additional constraint from LEP-1.5:

\parindent 0pt
{\bf 7.} The mass of the lighter chargino.  The unsuccessful direct LEP-1.5
searches for chargino pair production  mean that the chargino mass
must be greater than $67.8\, \hbox{GeV}$,
provided the chargino-LSP mass difference 
exceeds $10\, \hbox{GeV}$\cite{LEP1point5}. 
Direct searches for chargino pairs in 
the $Z^0$ decays at LEP-1 also require the lighter chargino mass to be above 
$45\, \hbox{GeV}$, consistent with constraint $1.\,\!$ above.
In fact, the inclusion
of the LEP-1.5 constraint renders constraint $1.$ mostly superfluous,
save in the narrow region of the parameter space where the 
chargino and the LSP are almost degenerate.  

\parindent 15pt

Many features of constraints $1.\!$-$3.$,$7.$ follow from the 
structure of the chargino mass matrix (see \cite{SUSY2} for 
its form and \cite{Pierce} regarding higher-order corrections not 
included here). The lighter chargino mass eigenvalue 
(absolute value) is lowered as $\tb$ increases; thus 
chargino mass limits for low $\mg$
tend to disallow large values of $\tb$.
Similarly, constraint $4.$ is dependent on the structure of the 
sfermion mixing matrices --- as $\mmu {\rm cot} \beta (\mmu \tb)$ 
increases (with $A_t = A_b = 0$), 
the lighter stop (sbottom) mass decreases. 
Constraints $5.$ and $6.$ above relate to the Higgs sector of the MSSM.
At tree level, the masses of the five Higgs bosons are fixed by 
inputting $\tb$ and the mass
of the $CP$-odd $\A$ \cite{Higrad}.
In general, consideration of the Higgs sector would introduce 
$\mA$ as another significant input
parameter which must be included in the analysis of the LEP data;
however, here we will restrict ourselves to the case in which 
$\mA {\sim}1\, \hbox{TeV}$.
In this case though $\mh$ can still be relatively light and 
constraint $6.$ can still rule out regions with 
$\tb$ close to unity and $\mmu$ less than 
$300\, \hbox{GeV}$ or so. 
If, on the other hand, one demands that $\A$ be quite light, 
the allowed parameter space is much more constrained.
 
Next consider briefly the effects of changing
the squark-sector input assumptions.
In SUGRA models
\cite{SUGRA}, a favored 
scenario is for $\mstl$ and $\mstr$ to be significantly smaller than
the other soft SUSY-breaking $\msq$'s (with the $\mstr$ also significantly 
smaller than $\mstl$).  We find that our results for high gluino masses
are not very sensitive to this change, since, as mentioned earlier, 
the presence of the top quark mass in the terms of the stop mass-squared 
matrix buoys up the physical stop masses for low values of $\mstl$ and 
$\mstr$.  In the pure MSSM, however,
{\em all} the soft SUSY-breaking squark masses are independent 
inputs \cite{FCNC}, and we find that lowering $\msbl$ and $\msbr$ 
below the common input for the first and second generation squarks 
($\msq$) {\em does} raise the LEP lower limit on $\msq$ significantly
for high gluino masses.  For low gluino masses, our results are sensitive 
to lowering the stop inputs though their effect on the $\h$ mass as
described below.
Non-zero $\aq$'s only significantly affect third generation squark masses 
and couplings since they appear
as $\mq \aq$, where $\mq$ is the mass of the relevant 
{\em quark}.  For the case of stops (and, to some extent, sbottoms)
we have verified that the effects of varying 
$\aq$ in the range --$2\, \hbox{TeV}$ to 
+$2\, \hbox{TeV}$ more or less duplicate those 
obtained by variation of $\mmu$ in the range $0$ to $1\, \hbox{TeV}$. 
This is as expected since the off-diagonal term in the stop
mass-squared matrix has the form $m_t (A_t - \mu {\rm cot} \beta)$. 
Thus variation of $\mmu$ to a large extent obviates the need to vary
$A_t$ and $A_b$.

In this present work, we wish to concentrate on constraints in the 
$\msq - \mg$ plane\cite{SUGsear},
allowing other input parameters to have any value in their
generally-accepted ranges, which we take to be:
$ 0 < \mmu < 1\, \hbox{TeV}$ and $1 < \tb < 35$ \cite{ftn1}.
The LEP constraints clearly disfavor light $\msq$'s or $\mg$'s
taken individually.  Further, squarks
alter the running gluino mass
leading to combined mass bounds rather than separate ones. 

Our results are shown in Figure 1(a) which illustrates bounds
from the seven constraints given above.
The shaded region bounded by solid lines
is ruled out; the remaining portion is allowed for {\it at least one value} 
of $\mu$ and $\tb$. The dashed lines show the bounds arrived at 
(for $\tb = 4$) from searches for squarks and gluinos  
by the CDF\cite{CDF,UA1} and D$\emptyset$\cite{D0} Collaborations. 
These trail off into big dots for the regions beyond the 
published limits. 
The small-dotted curve in Figure 1 illustrates the region excluded if 
the lower
bound on the chargino mass climbs to 85 GeV (provided the chargino-LSP
mass difference is larger than $10\, \hbox{GeV}$), 
which is more or less the discovery limit 
expected from 
LEP-2. The bounds on the squark-gluino mass plane
obtained and obtainable from LEP (again, with the gaugino unification
assumption) are seen to be somewhat complementary to 
those obtained from the Tevatron:
LEP covers more of the low $\mg$, high $\msq$ region than the Tevatron
while Tevatron does better in the moderate $\mg$, low $\msq$ 
region \cite{cvacua}.
LEP is also seen to exclude squark masses below about $70\, \hbox{GeV}$
for all gluino masses. 
Also, gluino masses much below $180\, \hbox{GeV}$ can be obtained only 
for large squark masses above $400\, \hbox{GeV}$ \cite{Wood}. 
Further note that if we add to this the projected LEP-2 bound that
 the mass of the chargino lie above 
$85\, \hbox{GeV}$, then a gluino mass below about 
$180\, \hbox{GeV}$ is
{\em ruled out irrespective of squark mass}.

Earlier studies
of the LEP \cite{LEPstud} (UA1 \cite{UA1}) constraints have yielded only 
$45$($53$)$\, \hbox{GeV}$ as a lower bound for the squark (gluino) mass.
Previous analyses (see for example figure 3 of \cite{Pierce} or 
Figure 6 of \cite{SUSY2}) also show a low $\mg$ window
for low $\tb$ and small negative values of $\mu$.
There is in fact a razor-thin band of 
$\mu$-choices (small in magnitude and negative)
for quite low gluino masses and for $\tb$ close to $1$ which
are allowed by the LEP-1 and LEP-1.5 constraints {\em on the 
chargino/neutralino sector}.  Here the coupling of the $Z^0$ to 
a lightest and a next-to-lightest neutralino is heavily suppressed.
In the low $\msq$ (and stop mass) `LEP-1' region of the figure this band 
is disallowed by constraint $6.$ on $\h$
(which in turn depends on the stop masses), and in the `LEP-2'
region the band is excluded by the chargino mass constaint \cite{hole}.

As alluded to above, Tevatron excludes a region of moderate gluino masses
($\mg \sim 300\, \hbox{GeV}$) and low squark masses 
($\msq \sim 100\, \hbox{GeV}$) which is allowed by the LEP constaints,
even with the unification hypothesis.  
It should be noted though that the
CDF and D$\emptyset$ analyses have only been presented for fixed values 
of $\tb$ and $\mu$ \cite{CDF,D0,musense,newrun},
and hence our analysis is not quite on a par with the assumptions 
going into their results. 
A more exact comparison is made in Figure 1(b), 
in which $\tb = 4$ as in \cite{CDF,D0}.  
Clearly, the extra region covered
by the Tevatron experiments which is not explorable at LEP persists.
LEP is also still seen to exclude squark masses below about 
$71\, \hbox{GeV}$ for all gluino masses. 
(This lower limit is unchanged for the case of $\mstl=\msbl=0.8\msq$,
$\mstr=0.6\msq$, but rises to roughly $93\, \hbox{GeV}$ for
$\mstl=\msbl=\msbr=0.5\msq$.)  In addition, a lower LEP bound of 
about $200\, \hbox{GeV}$ on the gluino mass holds for a much larger
range of squark masses with $\tb$ fixed at $4$ than the case shown
in Figure 1 where {\it all values} of $\tb$ are considered. 
In fact this bound appears to hold all the way up to a squark mass of 
$1.5\, \hbox{TeV}$ or more for $\tb = 4$. 
And for the `LEP-2' case the lower bound on gluino mass goes
as high as $260\, \hbox{GeV}$ for $\tb = 4$.

Finally, we wish to emphasize again that our results rely on the 
hypothesis of gaugino mass unification; if we give up this idea,
then the LEP constraints will have practically no effect on the squark
and gluino masses.  However, Tevatron data will still give constraints,
though not perhaps the same constraints as have been published, since 
these have also incorporated the gaugino mass unification assumption
into the analysis of the cascade decays of squarks and gluinos.

In this letter we have shown that gaugino mass unification 
and the running of the gluino mass enables LEP bounds on electroweak 
sparticle production to be translated into mass bounds 
on the strongly-interacting sparticles. These bounds depend
inseparably on both squark and gluino inputs and turn out to be 
comparable and, in some sense, complementary to those established at the 
Tevatron from direct searches for squarks and gluinos. Thus, studies 
of electroweak physics conducted at LEP can be a powerful
tools to probe some physics aspects normally thought to be accessible
only at a hadron collider. 

{\footnotesize
We thank N.K. Mondal and D.P. Roy for discussions and H. Baer and 
M. Drees for supplying some code modified for use in this work.
The work of MB is partially funded by a National Science Foundation
grant (No: 9417188). DKG wishes to acknowledge the University Grants 
Commission, India, for financial support. }


\begin{figure}[htb]
\epsffile[50 390 250 500]{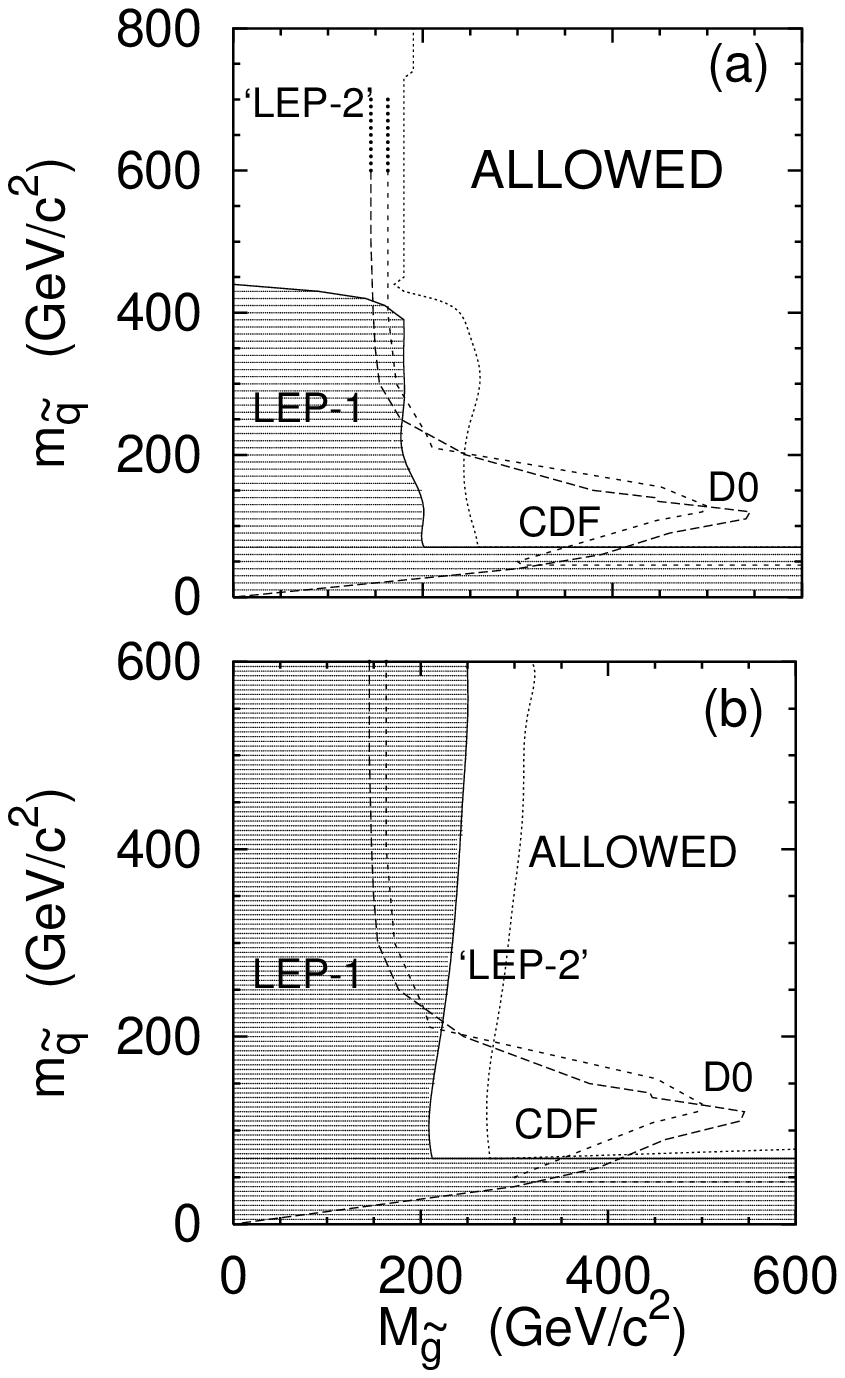} 
\vskip 320pt
\caption{
Illustrating the region in the $\msq - \mg$
plane allowed by LEP constraints. 
In (a) $\mu$ and $\tb$ are varied over their generally-accepted ranges,
while in (b) $\tb$ is fixed at $4$.
The shaded region is
ruled out by LEP-1 and LEP-1.5 constraints while the dotted 
curve delineates the 'LEP-2' region where the chargino 
mass is always less than 85 GeV. 
The dashed curves correspond to bounds established 
by the CDF (short dashes) and D$\emptyset$ (long dashes)
Collaborations for $\tb = 4$, trailing into big dots beyond the 
published results.
}
\end{figure}

\begin{thebibliography}{99}
\bibitem{scale} The selection of the machine energy as the scale at
which to evaluate the running mass is perhaps the most reasonable 
choice though by no means the only one possible.
\bibitem{degW1LSP} If the mass difference between the LSP and the next 
lightest neutralino is less than $3\, \hbox{GeV}$, then $Z^0$-decays
to such a pair will contribute to constraint $2.$ rather than
constraint $3.$
\bibitem{ftn1} Increasing these ranges has practically no effect on the
figure.
\bibitem{musense} The CDF analysis \cite{CDF} claims (see Figure 2
therein) that their result is insensitive to $\mu$.
\end{thebibliography}
\end{document}